\documentclass[a4paper,11pt]{article}
\usepackage{graphicx}
\usepackage{fullpage}
\usepackage{framed}

\usepackage{mystyle}
\graphicspath{{./}{figures/}}

\usepackage[table]{xcolor}
\usepackage{booktabs}
\usepackage{multirow}

\usepackage[displaymath, mathlines]{lineno}

\def\beq{\begin{eqnarray}}
\def\eeq{\end{eqnarray}}
\def\bea{\begin{eqnarray}}
\def\eea{\end{eqnarray}}
\newcommand{\ir}{\text{\tiny IR}}
\newcommand{\uv}{\text{\tiny UV}}

\usepackage{framed}
\usepackage{url}

\title{\sffamily
Warped Graviton Couplings to Bulk Vectors with Brane Localised Kinetic Terms}
\author{ Archil Kobakhidze$^1$, Kristian L. McDonald$^1$, Lei Wu$^{1,2}$ and Jason Yue$^1$ }

\affiliation[]{$^1$ ARC Centre of Excellence for Particle Physics at the Terascale, School of Physics, The University of Sydney, NSW 2006, Australia\\
$^2$ Department of Physics and Institute of Theoretical Physics, Nanjing Normal University, Nanjing, Jiangsu 210023, China
}

\abstract{{We realize non-universal couplings between Kaluza-Klein (KK) gravitons and bulk Standard Model (SM) vectors in the Randall-Sundrum (RS) model by including both UV and IR brane-localised gauge kinetic terms. We find that such kinetic terms can reduce the couplings of KK gravitons to SM gauge bosons and also ensure the KK vector masses are consistent with electroweak precision constraints.}\\[2cm]}

\begin{document}
\maketitle

\section{Introduction}

Multiple resonances naturally emerge in high-dimensional theories as the Kaluza-Klein (KK) excitations. Among high-dimensional theories, those based on warped anti-de Sitter (AdS) geometries \cite{Gogberashvili:1998vx, Randall:1999vf} are phenomenologically attractive since they predict various resonances at the $\sim$ TeV scale while ameliorating the hierarchy problem \cite{Randall:1999ee}. In the original RS model \cite{Randall:1999ee}, where only gravity propagates in the bulk, while Standard Model fields are comfined on the infrared (IR) brane, each KK graviton couples universally to the Standard Model fields and, therefore, it is has been tightly constrained by the negative results of searching for the high mass resonance in various channels \cite{CMS-PAS-EXO-14-005,Aad:2015kna,Aad:2015agg,Khachatryan:2015cwa,Aad:2014fha,Aad:2015mna,Khachatryan:2015yea,Aad:2015fna,CMS-PAS-B2G-12-006}  at the LHC.  For example, since ${\rm Br}(G_1\rightarrow l^+l^- )=0.5\times {\rm Br}(G_1\rightarrow \gamma\gamma)$ in the original RS model, the Run 1 dilepton searches constrain the lightest KK graviton $G_1$ to be heavier than $\sim 1$ TeV even for a small coupling $k/\overline{M}_p=0.01$ ($\overline{M}_p\approx 2.4\times10^{18}$ GeV is the reduced Planck mass).

Since the couplings of KK gravitons with other fields are essentially determined by the overlap of their wavefunctions in the extra-dimensional bulk spacetime, the phenomenology of the original RS model is altered when the Standard Model fields are allowed to propagate in the AdS bulk. Indeed, the coupling of $G_1$  to leptons get suppressed by localising them closer to the ultraviolet (UV) brane. The dilepton constraints, thus, can be avoided. However, in the bulk-RS model one necessarily encounters KK states for the Standard Model gauge bosons (gluons and electroweak bosons, collectively denoted by $A$), the lighest state of which, $A_1$, is, in general, expected to be lighter than $G_1$. Such setups generically generate sizable contributions to the electroweak precision observables and  corrections to the $Zb\bar{b}$ coupling \cite{Davoudiasl:1999tf}. It turns out that a smaller RS volume \cite{Davoudiasl:2008hx}, the introduction of a custodial symmetry \cite{Agashe:2003zs}, or large brane-localised terms \cite{Carena:2002dz} can allow for  new physics scales at $M_{A_1} \sim {2-5}$ TeV.

As is well known, the theoretical setup of the RS model involves non-local objects (3-branes) and thus allows one to write non-local operators localised on the IR and UV branes. In fact, even if such terms are absent at tree level, they are generically expected to appear radiatively. When allowed by the symmetries  of the theory, brane-localised operators that are quadratic in the  fields, such as brane localised kinetic terms (BLKTs)  and masses, can modify the mass spectrum of KK states, as well as their bulk profiles \cite{Carena:2002dz,Fichet:2013ola,Davoudiasl:2003zt,Davoudiasl:2002ua}. In this paper, we analyse the RS model with UV and IR brane localised terms  for both gravity and Standard Model (SM) gauge fields. Our interest is primarily in better exploring the sensitivity of the coupling between KK gravitons  and bulk vectors in the presence of BLKTs.

This paper is organised as follows. In Sec.~\ref{sec:2}, we calculate the bulk RS model with brane-localised kinetic terms. In Sec.~\ref{sec:3}  we present the numerical results and discussions. The paper will be concluded in Sec.~\ref{sec:5}.

\section{Bulk RS Model with Brane Localised Kinetic Terms}\label{sec:2}

\subsection{KK Graviton}

The RS model employs a five-dimensional spacetime with the following  invariant line-element:
\begin{equation}\label{eq:ds2}
  ds^2= e^{-2\sigma(y)} \eta_{\mu\nu} dx^\mu dx^\nu + dy^2:= G_{MN}dx^Mdx^N,
\end{equation}
where $x^{\mu,\nu}$ are the usual four-dimensional coordinates, $x^{M,N}$ are five-dimensional coordinates,  and  $y\in [-r_c\pi,r_c\pi]$ labels the extra dimension. The latter is compactified as an $S^1/\mathbb{Z}_2$ orbifold with fundamental interval $[0,\pi r_c]$, where $r_c$ denotes the radius of the extra space (we also use the notation $L=\pi r_c$ to denote the length of the fundamental interval). The exponential ``warp factor" depends on $\sigma(y):=ky$, with $k$ related to the AdS curvature; for numerical work we take $k$ in the vicinity of the four-dimensional reduced Planck mass  $k\sim \overline{M}_{p}=2.435\times10^{18}$ GeV.

In addition to the usual gravitational sources for the metric~\eqref{eq:ds2}, we augment the five-dimensional action to include four-dimensional brane curvature terms at both the Planck and TeV branes, located at $y = 0$ and $y=L$ respectively:
\begin{equation}
  S_G=\frac{M_5^3}{4}\int \sqrt{|G|} d^4x   dy \left[R^{(5)} + \left( g_{\uv} \delta (y) + g_{\ir} \delta (y -L) \right) R^{(4)} + \dots  \right].
  \label{eq:5d_action}
\end{equation}
The spectrum of spin-2 fluctuations (KK modes) is obtained by expanding the metric in terms of the fluctuation $h_{\mu\nu}$:
\begin{equation}\label{eq:met_pert}
  G_{\mu\nu} = e^{-2ky}(\eta_{\mu\nu} + 2M_5^{-3/2} h_{\mu\nu}),
\end{equation}
with the fluctuation further expanded in terms of the KK modes $h^{(n)}_{\mu\nu}$ as
\begin{equation}
  \begin{aligned}
    h_{\mu\nu}(x,y)& = \sum _{n} h^{(n)}_{\mu\nu} f^{(n)}_h(y).
  \end{aligned}
\end{equation}
Here $f^{(n)}_h(y)$ is the wavefunction for the $n$-th KK mode in the extra dimension, which must satisfy the equation of motion:
\begin{equation}\label{eq:KK_eom}
  \partial_y \left( e^{-4ky} \partial_y  f^{(n)}_h\right) + \left( 1+ \sum_{i=\textsc{\ir,\uv} }  g_{i} \delta(y-y_i)  \right) e^{-2ky}  m_n^2 f^{(n)}_h=0,
\end{equation}
where $y_{\uv,\ir} = 0,\,L$. The solutions are given in terms of the (modified) Bessel functions $J_i$ ($Y_i$) as:
\begin{equation}\label{eq:KK_wave_grav}
  \begin{aligned}
  f^{(n)}_h(y) &=\frac{e^{2ky}}{N_{n,h}} \left( J_2(z^G_n(y))+\alpha_n Y_2 (z^G_n(y)) \right),
   \end{aligned}
\end{equation}
where $ z^G_n(y):= m_{G_n}\frac{e^{ky}}{k}$ with $m_{G_n}$ being the KK mass.
Integrating (\ref{eq:KK_eom}) around the Planck and TeV branes gives the boundary conditions, enforcing which requires:
\begin{equation}\label{eq:roots_grav}
  \alpha_n = -\frac{ J_1(z_{n,i}^G)-\theta_i \gamma_i z_{n,i}^G J_2(z_{n,i}^G)}{Y_1(z_{n,i}^G)-\theta_i\gamma_i z_{n,i}^G Y_2(z_{n,i}^G)}\qquad i=\ir,\,\uv,
\end{equation}
where $z_{n,i}^G=z_n^G(y_i)$ and  $\theta_{\uv}=-\theta_{\ir}=-1$. Demanding that the two expressions for the constants $\alpha_n$ are equal determines the KK masses. The asymptotics for $\gamma_{\uv}: = g_0 kr_c/2 \neq -1/2$ gives $\alpha_n \sim z^2_n(0)\sim (10^{-15})^2$, and the mass of the low-lying KK modes is near the TeV scale. Thus, defining $x^G_n=z^G_n(L)$, the solutions for $x^G_n$ are well-approximated by solving:
\begin{equation}\label{eq:root_grav}
  J_1(x^G_n)+ \gamma_{\ir} x_n ^GJ_2(x^G_n)=0,
\end{equation}
giving the KK masses:
\begin{equation}\label{eq:KK_mass}
  m_{G_n} = x_n(\gamma_{\ir}) ke^{-kr_c \pi},
\end{equation}
where $x_n^G(\gamma_\ir)$ is the solution to Eq.~\eqref{eq:root_grav} for a given value of the parameter $\gamma_{\ir}:= g_{\ir} k r_c/2$. The normalisation factors $N_{n,h}$ are determined by the relation:
\begin{equation}\begin{aligned}
2\int_0^L dy\ e^{-2ky} \left[f_{h}^{(n)}\right]^2 + \sum_i \frac{2\pi \gamma_i}{kL} \left[f_{h}^{(n)}(y_i)\right]^2&=&1.
\end{aligned}\end{equation}

The 5D metric also contains a scalar fluctuation (the radion), which only acquires mass after the length of the extra dimension is stabilised. The properties of the radion are sensitive to brane-localised curvature and, in particular, the radion kinetic term, mass, and couplings all display dependencies on $\gamma_{\ir}$~\cite{George:2011sw,Dillon:2016bsb}. An important consequence is that the radion can be ghost-like when $\gamma_{\ir}>0.5$~\cite{George:2011sw,Dillon:2016bsb}. For values of $\gamma_{\ir}\sim10$ considered in this work (and related studies), one presumably requires additional dynamics to ensure stability of the background geometry (as discussed in Ref.~\cite{Dillon:2016bsb}). At present it is unclear whether the ghost-like radion  persists when the Higgs propagates in the bulk. For our purposes we merely point out that the RS geometry with IR curvature $\gamma_{\ir}\sim10$ requires an additional ingredient to ensure the radion is not ghost-like; optimistically,  a bulk non-minimal coupling to the Higgs may play a role here.\footnote{Ref.~\cite{Cox:2013rva} studied the effects of bulk SM fields on the radion couplings, however,  brane localised curvature terms were not considered.}

\subsection{Bulk Vector}

We consider the most-general action for a bulk vector:
\begin{equation}
\label{eq:5d_action_gauge}
    S= -\frac{1}{4 g_5^2}\int d^5 x \sqrt{-G}\, G^{MP} G^{NQ} F_{MN} F_{PQ}-\sum_{i=\ir,\uv}\frac{2\tau_{i}}{4 g_5^2k}\int d^4 x \sqrt{-G_i} \,G_i^{\mu \rho} G_i^{\nu\sigma} F_{\mu\nu} F_{\rho\sigma},
\end{equation}
where $G_i$ denotes the restriction of the 5D metric $G$ to the 4D induced metric on the brane at $y=y_i$, and $\tau_{\ir,\uv}$ are dimensionless coefficients for  brane-localised kinetic terms. The  KK expansion for the bulk vector is $A_\mu(x^\nu,y)=\sum_n A^{(n)}_\mu(x^\nu)\,f_A^{(n)}(y)$, where the profiles satisfy the following equation of motion:
\begin{equation}
  \begin{aligned}\label{eq:eom_gauge}
    \partial_y \left( e^{-2ky}\,\partial_y f^{(n)}_A(y)\right) +m_n^2\left(1+\frac{2}{k}\sum_i\tau_i \delta (y-y_i)\right) \, f^{(n)}_A(y)  = 0.
  \end{aligned}
\end{equation}
Integrating this equation over the boundaries gives the boundary conditions:
\begin{equation}\begin{aligned}
    \left ( k\partial_y  +\theta_i e^{2k y_i} m_{A_n}^2 \tau_i\right) \,f^{(n)}_A(y_i) = 0,
\end{aligned}\end{equation}
 The solutions are:
\begin{equation}\begin{aligned}
    f_{A}^{(n)}(y) &= \frac{e^{ky}}{\sqrt{2}N_{n,A}}\left\{ J_1 (z^A_n(y)) +\beta_n\,Y_1 (z^{A}_n(y))\right\},
\end{aligned}\end{equation}
where $z^A_n(y) =m_{A_n} e^{ky}/k$ and $N_{n,A}$ is a normalisation factor, and the constant $\beta_n$ is determined by the boundary conditions. Analogous to (\ref{eq:roots_grav}), the boundary conditions  on the branes give:
\cite{Davoudiasl:2002ua}:
\begin{equation}\label{eq:roots_gauge}
  \beta_n = -\frac{ J_0(z_{n,i}^A)-\theta_i \tau_i z_{n,i}^A J_1(z_{n,i}^A)}{Y_0(z_{n,i}^A)-\theta_i\tau_i z_{n,i}^A Y_1(z_{n,i}^A)},
\end{equation}
where $z^A_{n,i} := z^A_n(y_i)$. The vector KK masses $m_{A_n}$ are obtained after demanding equality of the two expressions for $\beta_n$. Due to $\lim_{z\rightarrow0}J_0(z) =1$, the mass of the first KK vector can be sensitive to both $\tau_{\uv}$ and  $\tau_{\ir}$. This differs from Eq.~(\ref{eq:root_grav}) where  $\lim_{z\rightarrow0}J_1(z) =0$, so that $m_{G_1}$ is only  sensitive to the values of the IR factor $\gamma_{\ir}$. The field renormalisation factor  are determined by the following relations:
\begin{equation}\begin{aligned}
    2k\int_0^L dy f^{(n)}_A(y)  f^{(m)}_A (y) +\sum_i 2\tau_i f^{(n)}_A(y_i) f^{(m)}_A(y_i)  = Z_n \delta_{mn,},
\end{aligned}\end{equation}
where
\begin{equation}\begin{aligned}
    Z_{n} &=  kL +2\sum_i\tau_i \left(f^{(n)}_A(y_i)\right)^2 {< kL +2\sum_i\tau_i \left(f^{(n>0)}_A(y_i)\right)^2} .
\end{aligned}\end{equation}

Using the above notation with $f^{(0)}_A(y)=1/\sqrt{2}$, one obtains a canonically normalised kinetic term for the zero-mode vector:
\begin{equation}\begin{aligned}
S&\supset& -\frac{1}{4 g_4^2}\int d^4 x\, \eta^{\mu\rho} \eta^{\nu\sigma} F^{(0)}_{\mu\nu} F^{(0)}_{\rho\sigma},
\end{aligned}\end{equation}
where the effective 4D coupling is
\begin{equation}\begin{aligned}
g_4^2 &=& \frac{g_5^2 k}{kL + \tau_{\uv}+\tau_{\ir}}.
\end{aligned}\end{equation}
This is the form of the coupling constant dictated by the AdS/CFT correspondence~\cite{Agashe:2002jx} (one can convert $kL$ to a logarithm to show the running of the CFT-induced contribution). Note that positivity of the gauge-coupling squared (equivalently, avoiding a ghost-like vector) requires $kL +\sum_i\tau_i>0$.

We wish to obtain the coupling of zero-mode vectors to KK gravitons. Labeling the SM vectors by the index $a=1,2,3$ for the $U(1)$, $SU(2)$  and $SU(3)$ factors, respectively,  and including the metric perturbation (\ref{eq:met_pert}) in the bulk vector action, gives
\begin{equation}\label{eq:grav_gauge}
  \begin{aligned}
    S\supset -\sum_{\substack{a\in\{ 1,2,3\}\\ n\in\mathbb{N}}}  \frac{c_{a,n}}{\Lambda}\int d^4 x \,h^{(n)}_{\mu\nu}\,T_{(0),a}^{\mu\nu},
  \end{aligned}
\end{equation}
where $ \Lambda= \overline{M_p} e^{-kL}$ is the warped-down IR cutoff scale, and the effective field-dependent couplings $c_{a,n}$ (here $a$ is the gauge index, and $n$ denotes the coupling to the $n$-th KK graviton) are defined through:
\begin{equation}\label{eq:ca_values}
  \begin{aligned}
    \frac{c_{a,n}}{\Lambda}&=& \frac{M_*^{-3/2}}
      {(kL+\sum_i\tau_{i}^a)}\left( kL\int_0^L\frac{dy}{L} f_h^{(n)}+\sum_i\tau^a_{i} f_{h}^{(n)}(y_i)\right).
  \end{aligned}
\end{equation}
Note that the index $a=1,2,3$ for the SM gauge factors appears because each gauge factor can have distinct UV and IR brane terms $\tau_{a,i}$. The zero-mode stress-energy tensor has the standard form:
\begin{equation}\begin{aligned}
T^{(0)}_{\alpha\beta}&=& \frac{1}{4}\eta_{\alpha\beta} F^{(0)}_{\mu\nu} F^{\mu\nu}_{(0)} -F^{(0)}_{\alpha\nu} (F^{(0)})^{\ \ \nu}_{\beta}.
\end{aligned}\end{equation}
One notes immediately that in the limit $\tau_{a,i}\rightarrow0$, the coupling of a given KK graviton to distinct vectors becomes universal, while for values of $\tau_{a,i}\ne0$, the couplings are, in general, non-universal. This differs from other bulk SM fields, for which the (in general) distinct bulk profiles are sufficient to generate non-universal couplings with the KK gravitons, via wavefunction overlap integrals. For the bulk vectors  (e.g. photon and gluon), the wavefunctions are identical when $\tau^a_i=0$, so nonzero BLKTs are a necessary ingredient to generate non-universal couplings. Note that, in the above, the zero-mode vectors have been rescaled, such that the gauge coupling $g_4$ now appears in the covariant derivative.

A further point of importance for this model relates to the localisation of SM fermions. While the lighter fermions are readily localised near the UV brane, as required to satisfy e.g.~dilepton constraints, the Higgs and top quark should preferably reside near the IR brane to address the hierarchy problem. However, this may be problematic if the KK1 graviton can decay preferentially to the top and Higgs. The presence of vector BLKTs can help here, which can enhance the coupling of KK gravitons to vectors, thereby making scenarios with IR localised top and Higgs consistent with the data. In our numerical analysis below, we focus on the case with UV localised Higgs and top quark for simplicity.


\section{Numerical results and discussions}\label{sec:3}

\begin{figure}[h!]\center
  \includegraphics[clip=true,trim = 5mm 2mm 15mm 15mm,width=.48\textwidth]{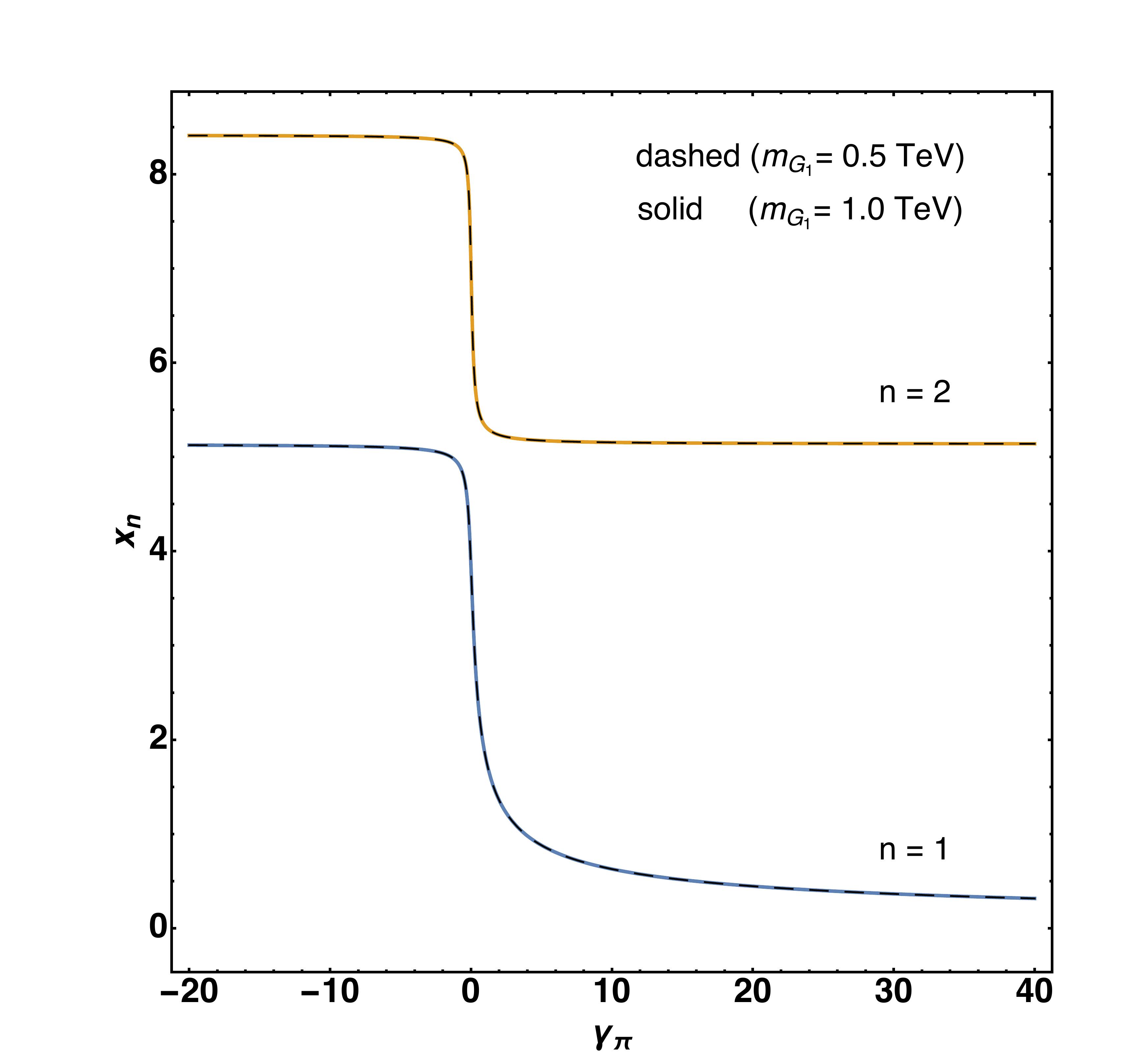}%
  \includegraphics[clip=true,trim = 5mm 2mm 15mm 15mm,width=.48\textwidth]{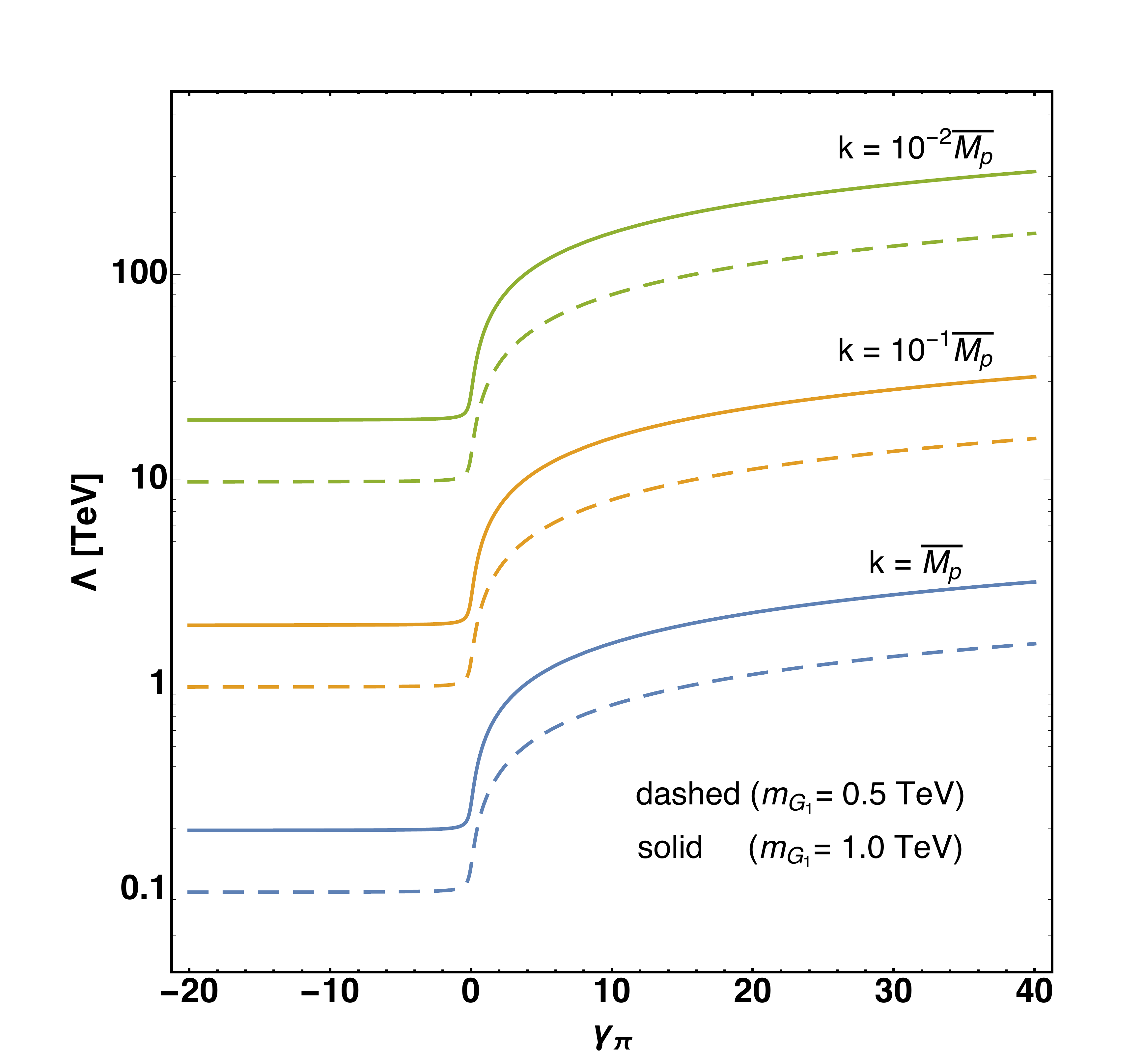}
  \caption{The first two roots $x^G_{1,2}$ of KK gravitons and the IR cut-off scale $\Lambda$ as functions of $\gamma_{\ir}$. }\label{fig:xn}
\end{figure}

In Fig.~\ref{fig:xn}, we present the first two roots $x^G_{1,2}$ of KK gravitons and the cut-off scale $\Lambda$ as functions of $\gamma_{\ir}$. To address the hierarchy problem, one typically requires values of $kr_c=\mathcal{O}(10)$. As mentioned previously, this gives $\alpha_n\sim 10^{-30}$, such that the low-lying KK graviton masses are largely insensitive to values of $\gamma_{\uv}\ne0$ (equivalently, the roots $x_n^G$ are essentially independent of $\gamma_{\uv}$). Thus, for simplicity, we assume $\gamma_{\uv}=0$ in our study. In the left panel of Fig.~\ref{fig:xn}, we see that $x_2^G$ is essentially constant away from the transition region near $\gamma_{\ir}=0$. However, for $\gamma_{\ir} > 0$, $x_1^G$ decreases like $\sim 2/\sqrt{\gamma_{\ir}}$ for increasing values of $\gamma_{\ir}$. In the right panel of Fig.~\ref{fig:xn}, we choose the value of the first KK graviton mass at $m_{G_1}=500~(1000)$ GeV and consider values of the curvature at $k=(10^{-2},10^{-1},1) \times \overline{M}_{p}$. With $m_{G_1}=x_1^G ke^{-kr_c\pi}$, one can determine the radius of compactification $r_c$ for each given $k$ value. For a given fixed value of $\gamma_{\ir}$, smaller values of $k$ require smaller values of $r_c$. The radius $r_c$ affects the IR scale $\Lambda$ via the warp factor $e^{-kr_c\pi}$. It can be seen that the IR scale $\Lambda$ can be significantly lifted by lowering the value of $k$. When $\gamma_{\ir} > 0$, the value of $\Lambda$ increases rapidly for increasing values of $\gamma_{\ir}$. To obtain $\Lambda \gtrsim $ 1 TeV, the value of $\gamma_{\ir}$ needs to be larger than about 10 (20). In the studies that follow, we take $k=\overline{M}_{p}$ and $m_{G_1}=500~(1000)$ GeV as examples.

\begin{figure}[!h]\center
  \includegraphics[clip=true,trim = 75mm 20mm 30mm 20mm,width=0.475\textwidth]{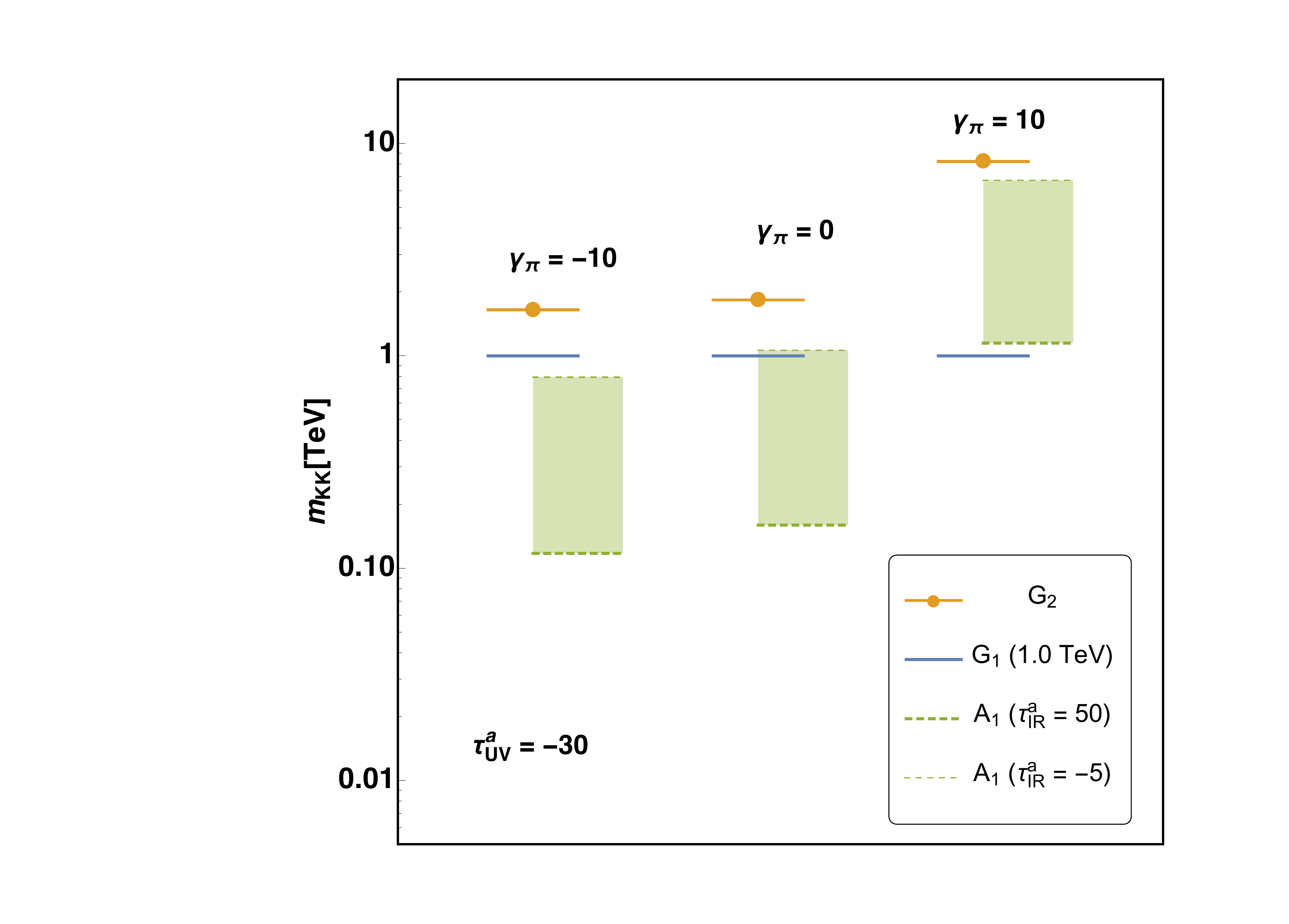}%
  \includegraphics[clip=true,trim = 75mm 20mm 30mm 20mm,width=0.475\textwidth]{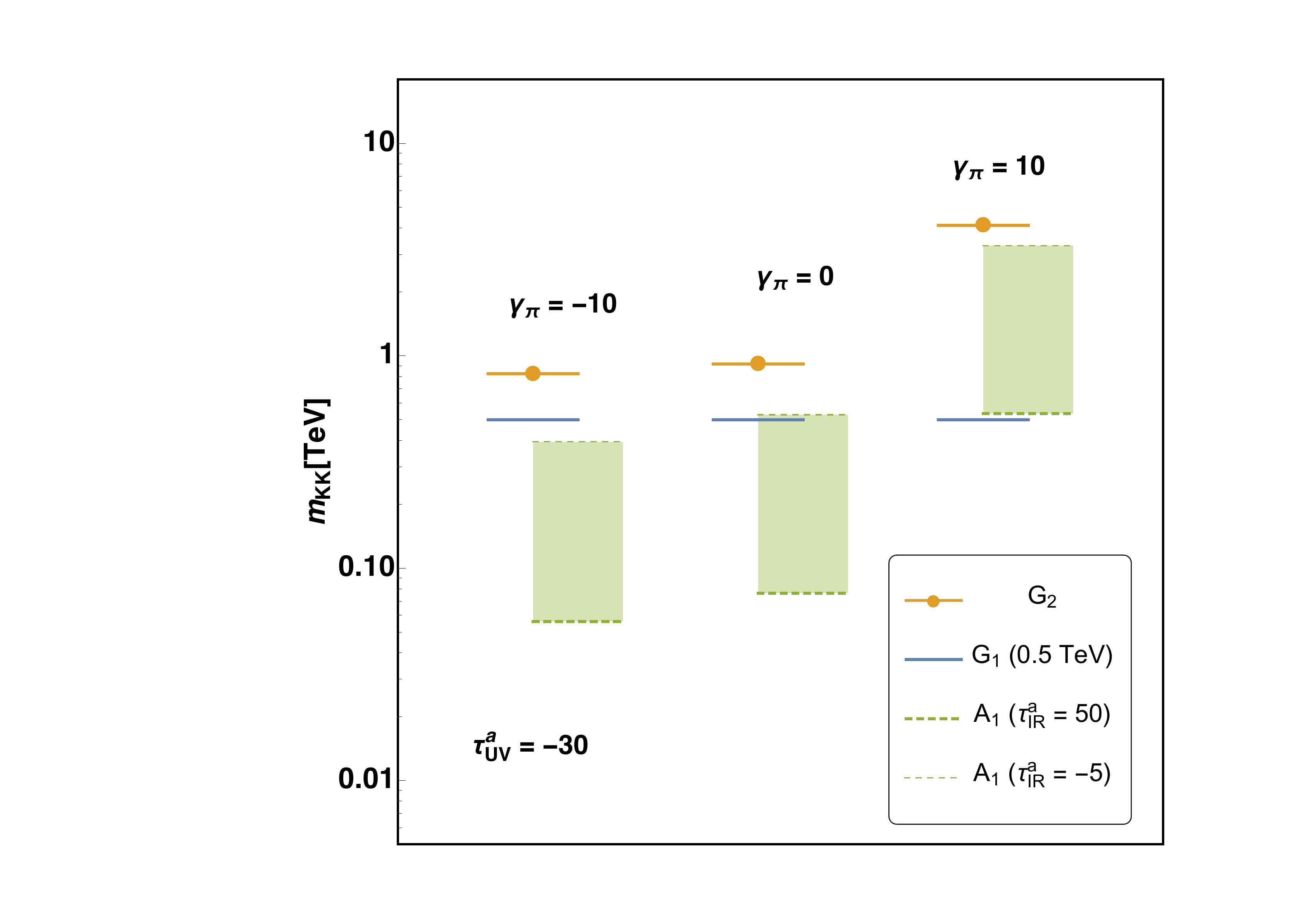}%
  \caption{Masses of the first two KK gravitons $G_{1,2}$ and the first KK vector bosons $A^a_1$ for $\tau^a_{\uv}=-30$ with $\tau^a_{\ir}$ varying in the range $[-5, 50]$, where $a\in\{1,2,3\}$. \label{fig:spectrum}}
 \end{figure}

It should be noted that the warped down scale can not be too low. Otherwise, the KK vector masses are sufficiently light to induce tension with precision constraints. In Fig.~\ref{fig:spectrum}, we present the mass spectrum of the first two KK gravitons and the first KK vector bosons by including the BLKT terms $\tau_{\uv}$ and $\tau_{\ir}$. Here we assume $\tau_{\uv}=-30$ and vary $\tau_{\ir}$ in the range $[-5, 50]$ to examine the effect of the IR BLKT  on the KK vector masses. The lower limit of $\tau_{\ir}$ results from the requirement that $\tau_{\uv}+\tau_{\ir}+k \pi r_c >0$ in order to avoid ghost-like vectors. With $m_{G_1}=500~(1000)$ GeV and $k=\overline{M}_{p}$, the second KK graviton mass $m_{G_2}$ is determined by $\gamma_{\ir}$ via the ratio of the first two roots $x^G_2/x_1^G$. This ratio is nearly constant in the negative $\gamma_{\ir}$ region and grows substantially large in the positive $\gamma_{\ir}$ region. Thus, $m_{G_2}$ is predicted to be around 0.8 (1.7) TeV for $\gamma_{\pi}<0$ but goes up to about 4 (8) TeV for $\gamma_{\ir}=10$. On the other hand, the KK vector masses are determined by both $\gamma_{\ir}$ (through the corresponding changes in $r_c$) and the BLKTs $\tau_{\uv,\ir}$. For $\gamma_{\ir} \lesssim 10$, as a result of the low IR scale, the KK vector masses are basically less than $m_{G_1}$, which is disfavored by the eletroweak precision measurements. However, we can see that a negative $\tau_{\ir}$ can significantly lift the KK vector boson masses, while the positive $\tau_{\ir}$ can reduce them. For example, when $\gamma_{\ir}=10$ and $m_{G_1}=1$ TeV, the KK vector masses can reach about 7 (1.2) TeV for $\tau^a_{\ir}=-5 (50)$.


\begin{figure}[h!]\center
  \includegraphics[clip=true,trim = 15mm 2mm 5mm 10mm,width=.475\textwidth]{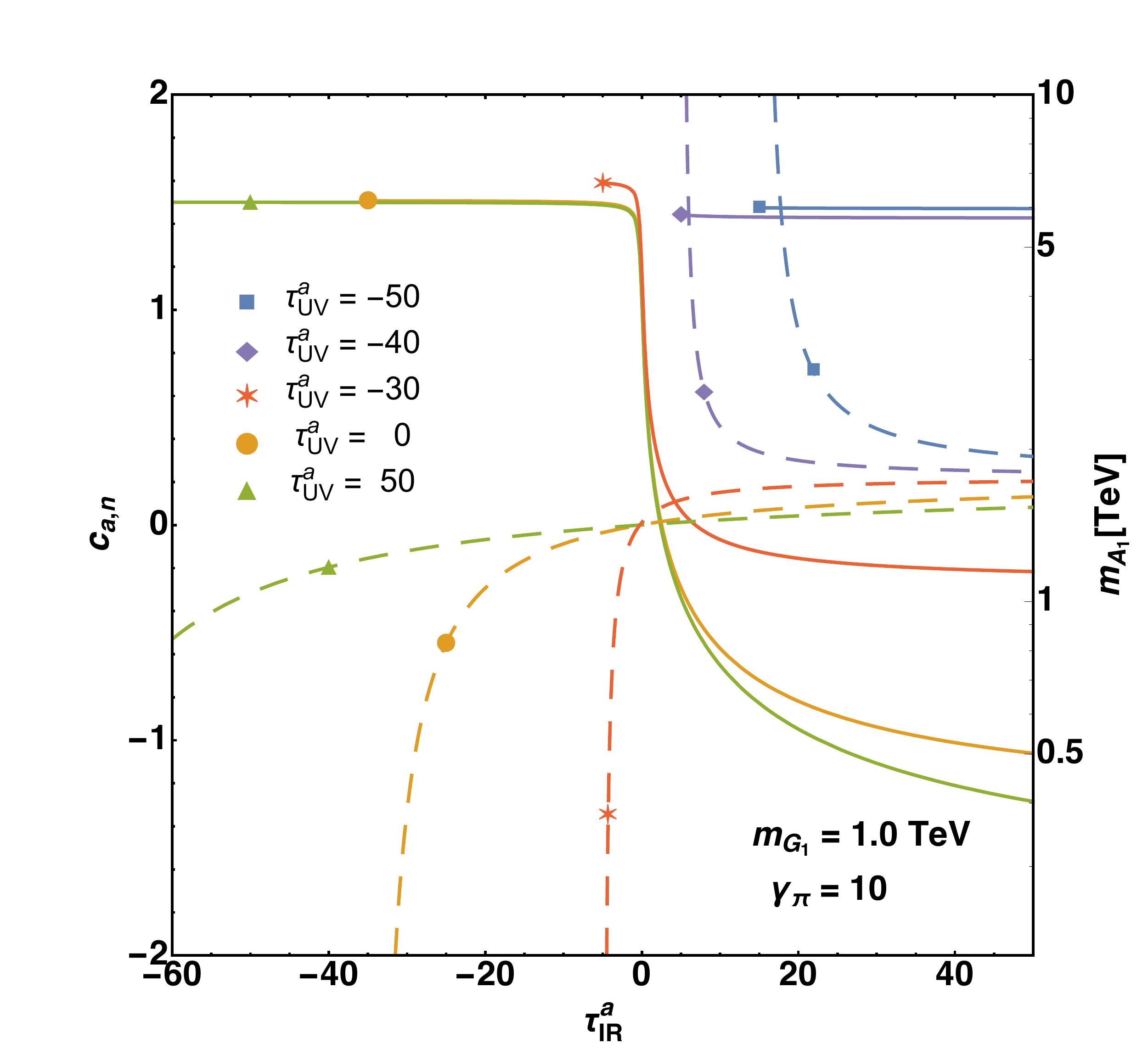}%
  \includegraphics[clip=true,trim = 15mm 2mm 5mm 10mm,width=.475\textwidth]{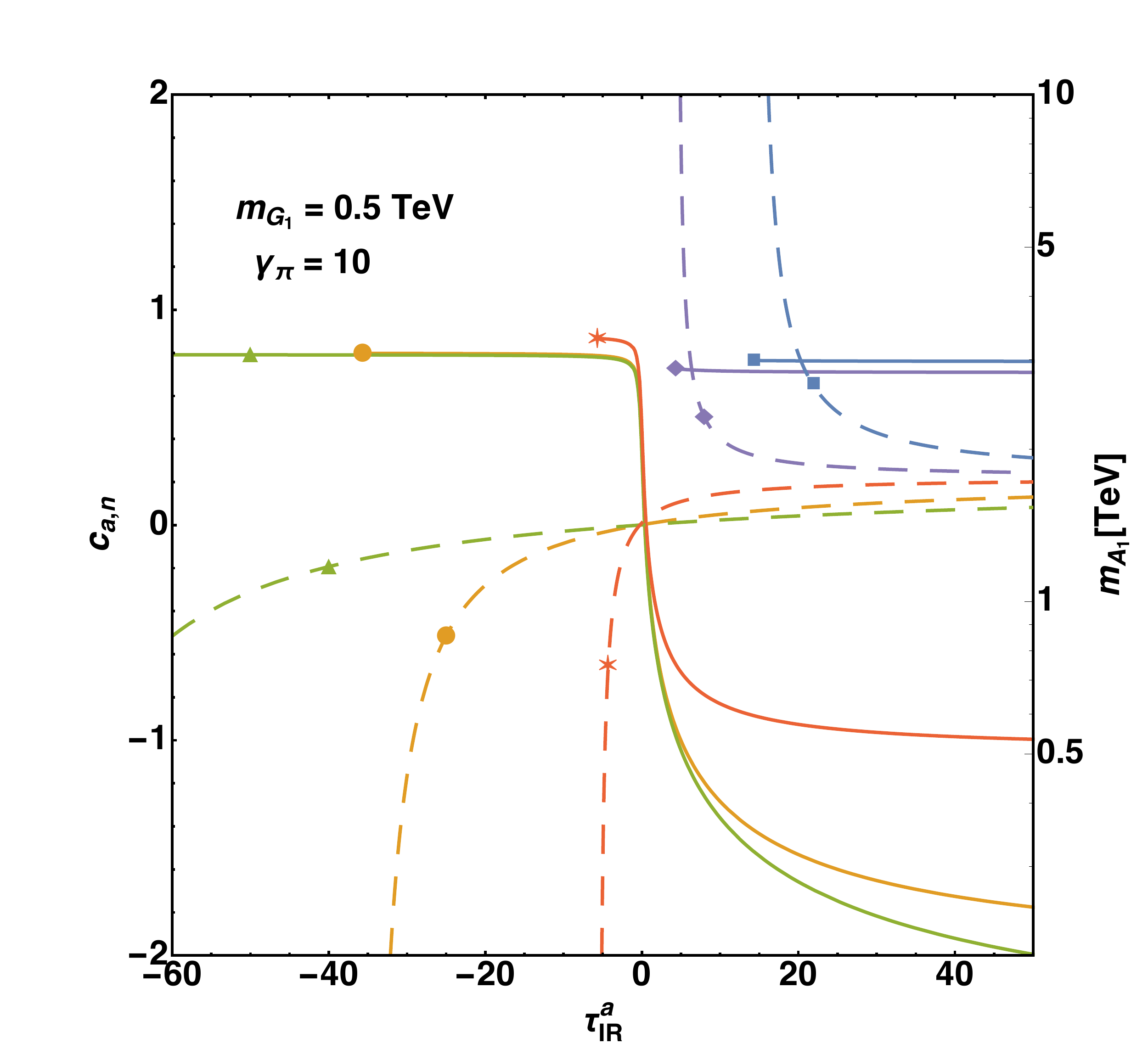}%
  \caption{The gauge coupling $c_{a,1}$ of KK graviton $G_1$ as a function of $\tau^a_{\ir}$ for $\tau^a_{\uv}\in \{ -50,-40,-30,0,50\} $, where the gauge index $a\in\{ 1,2,3\} $. The dashed and solid lines correspond to the gauge couplings $c_{a,1}$ and KK vector boson mass $m_{A_1}$, respectively. The condition $\tau_{\uv}+\tau_{\ir}+k \pi r_c >0$ is imposed to avoid ghost-like vectors. \label{fig:ca_mA} }
  \vspace{-0.5cm}
 \end{figure}
To reproduce the diphoton signal one must also consider the couplings $c_{a,n}$ between the zero-mode vector of type $a$ and the $n$-th KK graviton. In Fig.~\ref{fig:ca_mA}, we plot the coupling $c_{a,1}$ of the KK graviton $G_1$ as a function of $\tau^a_{\ir}$ for $\tau^a_{\uv}=-50,-40,-30,0,50$, where the gauge index $a=1,2,3$. The condition $\tau_{\uv}+\tau_{\ir}+k \pi r_c >0$ is imposed to avoid ghost-like vectors. We can see that when $\tau^a_{\ir}>0$, the coupling $c_{a,1}$ can be enhanced by increasing $\tau^a_{\ir}$, however, the corresponding KK vector mass $m_{A,1}$ will be reduced significantly for $\tau^a_{\uv}=-30,0,50$. This may cause tension between the diphoton enhancement and the electroweak precision observables. But for $\tau^a_{\uv}=-40,-50$, the coupling $c_{a,1}$ decreases with increasing $\tau^a_{\ir}$ and the corresponding KK vector mass $m_{A,1}$ will be around 4 TeV and keep almost constant. On the other hand, when $\tau^a_{\ir}<0$, the sign of coupling $c_{a,1}$ will be changed to negative and decreases if $\tau^a_{\ir}$ becomes smaller.

Finally, we mention that our analysis corresponds to the case where all other SM fields are localised towards the UV brane. However, while the light fermions can, in general, be localised towards the UV brane, ideally one would like the Higgs and top quark to be localised near the IR brane, to address the hierarchy problem. Using an IR BLKT for the bulk hypercharge field is able to render the model consistent with an IR localised Higgs boson and top quark, such as in the limit with only $\tau^1_{\ir}\ne0$. Our results also show that one can  obtain additional regions of parameter space consistent with an IR localised Higgs and top quark by taking $\tau^1_{\uv}\ne0$.



\section{Conclusions}\label{sec:5}

In this work we studied the RS model with brane-curvature and the most-general set of BLKTs for bulk vectors. Unlike bulk fermions and scalars, for which different bulk wave functions are sufficient to generate non-universal couplings to KK gravitons, even in the absence of brane-localised terms, massless bulk vectors typically have (identical) flat profiles and thus couple universally to KK gravitons. However, the use of BLKTs modifies this expectation and allows non-universal couplings between bulk vectors and KK gravitons. We showed that the most-general set of vector BLKTs allows one to enhance the coupling between the KK graviton and SM vectors. In particular, we showed that UV BLKTs allow regions with larger couplings, while permitting KK vector masses of $m_{A_1}\gtrsim 2$~TeV, so that both IR and UV BLKTs can play an important role in such models, supporting the scenarios with IR localised Higgs and top quark.

\acknowledgments
This work is partially supported by the Australian Research Council. AK was also supported in part by the Shota Rustaveli National Science Foundation (grant DI/12/6- 200/13). LW was also supported in part by the National Natural Science Foundation of China (NNSFC) under grants Nos. 11705093, 11305049.


\bibliographystyle{mybibsty}
\bibliography{myrefs}
\end{document}